\documentclass[aip,jcp,twocolumn,superscriptaddress,reprint]{revtex4-1}
\usepackage{graphicx}
\usepackage{amsmath}
\usepackage[usenames, dvipsnames]{color}
\usepackage[normalem]{ulem}

\newcommand{\add}{\color{black}}
\newcommand{\ts}{\textsection}
\newcommand{\be}{\begin{equation}}
\newcommand{\ee}{\end{equation}}
\newcommand{\ltsim}{\protect\raisebox{-0.5ex}{$\:\stackrel{\textstyle <}{\sim}\:$}}

\newcommand{\f}{\frac}
\newcommand{\md}{\mathrm{d}}

\def\deg{$^{\circ}$}
\newcommand{\kT}{k_{\rm B}T}

\newcommand{\biR}{\mathbf{R}}
\newcommand{\avCos}{\langle\cos\theta\rangle}
\newcommand{\thetaJoin}{\theta_{\rm join}}
\newcommand{\cosThetaJoin}{\cos\thetaJoin}
\newcommand{\Dm}{\Delta\mu}
\newcommand{\bDm}{\beta\Dm}
\newcommand{\ODm}{\overline{\Dm}}
\newcommand{\bODm}{\beta\ODm}
\newcommand{\lP}{\ell_{\rm P}}

\newcommand{\lPH}{\lP^{\rm H}}
\newcommand{\lPM}{\lP^{\rm M}}

\begin{document}
\preprint{}
\title{Consequences of local inter-strand dehybridization for large-amplitude bending fluctuations of double-stranded DNA}

\author{David A. Sivak}
\altaffiliation[Current address: ]{Physical Biosciences Division, Lawrence Berkeley National Laboratory}
\author{Phillip L. Geissler}
\email[]{geissler@berkeley.edu}
\altaffiliation[]{Department of Chemistry, University of California, Berkeley}
\altaffiliation[]{Chemical Sciences Division, Lawrence Berkeley National Laboratory}
\affiliation{Biophysics Graduate Group, University of California, Berkeley}

\begin{abstract}
The wormlike chain (WLC) model of DNA bending accurately reproduces single-molecule force-extension profiles of long (kilobase) chains. These bending statistics over large scales do not, however, establish a unique microscopic model for elasticity at the 1-10 bp scale, which holds particular interest in biological contexts. Here we examine a 
{\add class of microscopic models} which allow for disruption of base pairing (i.e., a `melt' {\add or `kink', generically an `excitation'}) and consequently enhanced local flexibility. We first {\add analyze the effect on the excitation free energy of integrating out the spatial degrees of freedom in a wormlike chain. Based on this analysis, we present a formulation of these models that ensures} consistency with the well-established thermodynamics of melting in long chains. {\add Using a new method to calculate cyclization statistics of short chains from enhanced-sampling Monte Carlo simulations}, we compute {\add $J$-factors} of a meltable wormlike chain (MWLC) over a broad range of chain lengths, including very short molecules (30 bp) that have not yet been explored experimentally. For chains longer than about 120 bp, including most molecules studied to date in the laboratory, we find that melting excitations have little impact on cyclization kinetics. Strong signatures of melting, which might be resolved within typical experimental scatter, emerge only for shorter chains.
\end{abstract}

\date{\today}
\maketitle

\section{Introduction \label{sec:intro}}
{\add The challenge of constructing coarse-grained models, e.g., for large biomolecular systems, lies in accounting for the influence of fluctuations that are not explicitly represented. Systematic procedures for coarse-graining perform such an accounting, often approximately, so that physical consequences of fluctuating solvent densities, electric fields, etc. can be incorporated at low computational cost. In the simplest cases, integrating out certain degrees of freedom just renormalizes interaction parameters for the remaining variables. In general new types of interactions are introduced as well.

This familiar process of renormalization has important implications for the way experimental data should be used to parameterize microscopic models. Measured statistics of a particular variable include the effects of all other fluctuating degrees of freedom. Thermodynamic parameters inferred for that variable do not directly inform the energetics of models that explicitly include other degrees of freedom. In some situations these considerations are transparent: an experimentally derived implicit solvent potential should not be used in models that explicitly represent solvent fluctuations. In other cases the issue can be somewhat more subtle. This paper demonstrates the importance of disentangling renormalization effects in the specific context of models for bending of nucleic acid molecules.}

The wormlike chain (WLC) model has proven remarkably successful in reproducing experiments probing the conformational flexibility of DNA.  In a discretized form the model envisions DNA as a fluctuating chain of discrete links, each inextensible in its length \cite{Bustamante:2000p5920,Smith:1996p21972}, with the persistence length $\lP\sim50$ nm ($\sim$150 basepairs [bp]) setting the contour length scale over which orientational correlations decay. The model shows very good agreement with single-molecule force-extension measurements on kilobase-long $\lambda$-phage genomic DNA.~\cite{Bustamante:2000p5920,Bustamante:2003p5816} The WLC with $\lP=50$ nm also accurately predicts cyclization rates of medium-length (hundreds of bp) DNA chains in ligation experiments.~\cite{Crothers:1992} {\add In general the WLC model reproduces the results of experiments whose observations are dominated by conformations typical of thermal equilibrium.}

Yet in biological contexts DNA is often bent on much shorter length scales than the micron lengths of $\lambda$-phage DNA: prokaryotic transcription initiation,~\cite{PerezMartin:1997p21677} nucleosomal genome compaction in eukaryotes,~\cite{Richmond:2003p7229} DNA-binding by architectural proteins such as IHF,~\cite{Rice:1996p7683} and viral DNA packaging~\cite{Garcia:2007p15000} all feature DNA bending hundreds of degrees on length scales of tens of bp. Furthermore, DNA is increasingly being used as a programmable template for constructing nanomaterials,~\cite{Mirkin:1996p3649,Fu:2004p32999,Claridge:2008p3392,Dietz:2009p41617} where its bending flexibility over tens of bp can strongly influence the resulting structures.~\cite{Park:2008p16914} Knowledge of the mechanical properties of short DNA chains is thus essential to understand the role of DNA looping in gene regulation, the nature of protein-DNA interactions, the pressure generated in viral DNA packaging, and the patterned nanomaterial dictated by a given DNA scaffold. 

Many microscopic bending potentials produce long length scale bending statistics identical to the WLC,~\cite{Wiggins:2006p22097} and thus the relatively well-established long length scale results do not distinguish between substantially different possibilities for smaller length scale bending potentials. In particular, the energetics associated with large end-to-end forces, or equivalently, during exceedingly rare large thermal fluctuations, could take on numerous different forms without substantially changing the long length scale bending statistics. 

Indeed, recent experiments have suggested that oligomers shorter than a single persistence length may have considerably different mechanics than a WLC with $\lP=50$ nm. Measuring the end-to-end distance of DNA free in solution by F\"{o}rster Resonance Energy Transfer (FRET) and its radius of gyration by small angle X-ray scattering (SAXS), Yuan, \emph{et al.} have inferred an {\add apparent} persistence length of $20$ nm for chains comprising fewer than $21$ bp,~\cite{Yuan:2008p8870} though it is not clear how this conclusion is consistent with the accepted value inferred from experiments on longer DNA chains. {\add Cyclization rates of short chains are particularly dominated by rare large fluctuations, and thus should provide a sensitive test of DNA's detailed short length scale bending statistics. Indeed,} DNA ligation experiments by Cloutier and Widom found that DNA of approximately 100 bp cyclized as much as five orders of magnitude more readily than expected from corresponding WLCs with $\lP=50$ nm.~\cite{Cloutier:2004p4086} By contrast, in similar experiments Du, \emph{et al.} found the cyclization efficiency of small DNA molecules to be in good agreement with the traditional WLC model of DNA bending albeit with a slightly shorter persistence length ($\lP$, 47 nm).~\cite{Du:2005p1748}

It is easy to imagine that the collective deformations of a uniformly buckled rod, when subjected to extreme bending, could be superseded by localized excitations that render short stretches of the chain very pliable. Recent theoretical work has sought to account in this manner for the anomalous rates of Cloutier and Widom {\add by positing a wormlike chain model incorporating such thermally-excited `melts' or `kinks'. The apparent persistence length (after integrating over the excitations) of such a model is greater than that of the conventional wormlike chain: melting influences bending. In this work we demonstrate that, conversely, it follows that bending influences melting: the apparent free energy of forming such an excitation (after integrating over the spatial degrees of freedom) is less than the `bare' free energy directly entering into the model Hamiltonian.} In \ts\ref{sec:graft} we show that as originally parameterized, the {\add meltable wormlike chain} theory of Yan and Marko~\cite{Yan:2004p1403,Yan:2005p1413} improperly accounts for {\add this} entropic stabilization of {\add enhanced-flexibility excitations}. Specifically, with the original thermodynamic penalty for local melting, that model appears to be inconsistent with well-characterized thermodynamics of bulk DNA melting. We discuss this double-counting of certain entropic gains due to melting, and {\add in \ts\ref{sec:param}} propose a revised parameterization for the `meltable' WLC model. {\add In \ts\ref{sec:sims} we outline a new method for evaluating $J$-factors that characterize cyclization kinetics, from enhanced-sampling Monte Carlo (MC) simulations. This method could be applied to any comparably coarse-grained model and, unlike existing methods, incurs similar computational cost for both long and short DNA chains. In \ts\ref{sec:jF} we explore} cyclization rates using {\add these} MC simulations, {\add finding} that such excitations, properly parameterized by well-established DNA melting thermodynamics, produce no practical difference in cyclization rates for DNA chains longer than $\sim$120 bp. Only for shorter chains do models including melting excitations produce significantly higher cyclization rates than the unmeltable WLC model, though this rate enhancement is lower than that seen in some experiments.

\section{The wormlike chain model with thermal excitations of enhanced flexibility \label{sec:modelMWLC}}
\begin{figure*}[!htbp]
\centering
\includegraphics{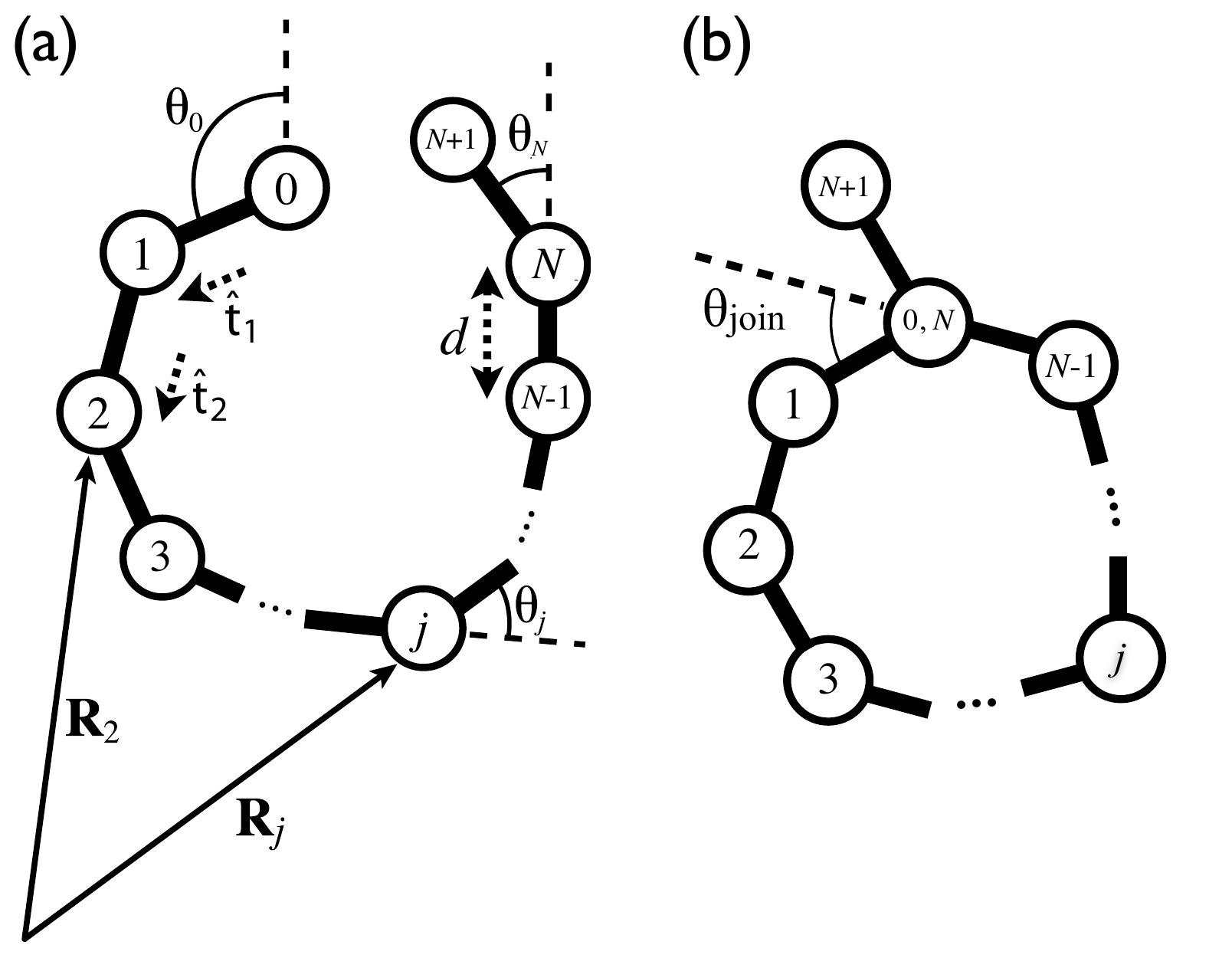}
\caption{Geometry of a meltable WLC comprising $N+1$ links.  (a) Chain cyclization amounts to imposing the constraints $\biR_{0}=\biR_N$, $\theta_0=\theta_N$, and $\phi_0 = \phi_N$; where $\phi_0$ and $\phi_N$ are azimuthal angles (not shown) corresponding to the polar angles $\theta_0$ and $\theta_N$, respectively. Note that the angle $\theta_0$ is defined with reference to the vector $\biR_N - \biR_{N-1}$, which is oriented vertically in this example configuration.  (b) To determine the propensity of apposed chain ends to attain similar relative orientation, biased simulations (see \ts\ref{sec:relatingJ}) calculate the conditional probability distribution of $\cosThetaJoin$, given $\biR_N=\biR_0$, where $\theta_{\rm join}$ is the angle between the vectors $\biR_N-\biR_{N-1}$ and $\biR_1-\biR_0$.} \label{fig:chain}
\end{figure*}

The WLC model envisions DNA as a single chain of uniform-length segments (each representing a fixed number of basepairs) connected at a set of nodes, where the bend angle at a given node $i$ is allowed to fluctuate according to a bending energy quadratic in the local curvature
\begin{subequations}
\begin{align}
\beta E_i^{\rm {\add WLC}} &= \f{1}{2}\f{\lP}{d}|\hat{t}_{i+1} - \hat{t}_i|^2\\
&= \f{\lP}{d} (1-\cos\theta_i)\ .
\end{align}
\end{subequations}
Here $\beta=(\kT)^{-1}$, $\lP$ is the persistence length, $\hat{t}_i$ is the unit-length vector pointing from node $i-1$ to node $i$, $d$ is the contour length separating adjacent nodes, and $\theta_i = \cos^{-1}(\hat{t}_{i+1}\cdot\hat{t}_i)$ is the local bending angle (Fig.~\ref{fig:chain}a). 

The meltable WLC (due to Yan and Marko~\cite{Yan:2004p1403,Yan:2005p1413}) adds an additional set of variables, the hybridization states of those nodes: each node has two internal states, notionally corresponding to hybridized and melted local structures. Fluctuations in base pairing in this model are represented by changes in the hybridization state. The free energy associated with a single node in hybridization state $m$ and bent at an angle $\theta$ is
\begin{align}
\beta E^{\rm {\add MWLC}}(m,\theta) &= \delta_{m,0}\left[\f{\lPH}{d}(1-\cos\theta)\right] \label{equ:mwlc}
\\
&\ \ \ \ \ + \delta_{m,1}\left[\bDm(T) + \f{\lPM}{d}(1-\cos\theta)\right]. \notag
\end{align}
where $\lPH$ is the persistence length of hybridized DNA and $\lPM$ is the persistence length of molten DNA. We have defined $m=0$ as the hybridized state and $m=1$ as the molten state. The Kronecker delta, $\delta_{i,j}$, is 0 if $i\ne j$ and 1 if $i=j$. The thermodynamic penalty for melting $\Dm$ is the reversible work required to disrupt base pairing at a node of the chain that has a fixed bend angle $\theta = 0$.

{\add We also compare the MWLC with two other models originally formulated from similar perspectives. The kinkable wormlike chain (KWLC)~\cite{Wiggins:2005p1466} posits a completely flexible molten state, $\lPM=0$. The spontaneous bend model maintains some rigidity in the molten state but introduces a preferred bend angle $\cos^{-1}\gamma$:
\begin{align}
\beta E^{\rm SB}(m,\theta) &= \delta_{m,0}\left[\f{\lPH}{d}(1-\cos\theta)\right] \label{equ:ib}
\\
&\ \ \ \ \ + \delta_{m,1}\left[\bDm(T) + \f{\lPM}{d}(\cos\theta - \gamma)^2\right]. \notag
\end{align}
Nevertheless, the specification of a preferred bend angle seems less physically plausible for bare DNA than for protein-bound DNA, and the assumption of completely flexible molten regions is unnecessarily extreme for our purposes. Thus we focus primarily on the role of thermally-excited local regions of enhanced flexibility (`melts') in the cyclization kinetics of DNA.}

In {\add these models} a chain of contour length $(N+1)d$ has $N+2$ nodes (which we number $0$ to $N+1$) connected by $N+1$ fixed-length links, and $N$ bending degrees of freedom (Fig.~\ref{fig:chain}a). Lacking bending energy, the first and last nodes can be ignored in writing the chain's total energy:
\be
E_{\rm chain}[\{m_i\}, \{\theta_i \}] = \sum_{i=1}^N E(m_i,\cos\theta_i). 
\label{equ:wlce}
\ee
Because nodes contribute to $E_{\rm chain}$ in an additive fashion, they fluctuate in a statistically independent manner (in the absence of collective constraints). 

In considering a model chain that is inextensible along its contour, we neglect potential effects of stretching fluctuations on cyclization. This approximation is supported by estimates of the dsDNA stretch modulus, $\sim$1000 pN,~\cite{Smith:1996p21972,Wang:1997p1833} and by the threshold force, $\sim$65 pN, for the overstretching transition to a form $\sim$1.7$\times$ longer.~\cite{Bustamante:2003p5816} Both values well exceed the typical range of forces anticipated in the course of loop closure. This model only implicitly accounts for structural aspects on length scales smaller than a single link, such as the detailed atomic structure of the DNA basepair. 

\section{Cyclization kinetics \label{sec:cycKin}}
For decades, cyclization reactions have been a popular experimental tool for measuring DNA bending elasticity.~\cite{Crothers:1992} A population of DNA chains of given length and sequence, with `sticky' single-stranded overhangs on each end, are incubated in solution with a DNA ligase, which can covalently link the two ends together when they come into spatial proximity and appropriate relative orientation. Samples are taken at periodic time intervals and run on a gel to quantify the fluorescence from bands corresponding to populations of circularized monomers, linear dimers, and so on. Dividing the rate constant for cyclization $k_{\rm cyc} \equiv \lim_{t\rightarrow 0}\, [\md C(t)/\md t] \, / \, M_0$ by the rate constant for dimerization $k_{\rm dim} \equiv \lim_{t\rightarrow 0} \, [\md D(t) / \md t] \, / \, M_0^2$, controls in certain limits for the ligase enzyme's propensity to ligate apposed sticky ends, yielding a measure of chain flexibility, the so-called $J$-factor, $J = 2k_{\rm cyc}/k_{\rm dim}$. Here $C(t)$ and $D(t)$ are the concentrations of ligated circular monomers and ligated linear dimers, respectively, and $M_0$ is the initial concentration of unligated monomers.   

In the case of low ligase concentration, ligation should proceed at a rate proportional to the equilibrium population of precursors adopting ligation-competent conformations. We define a monomer conformation as ligation-competent when one end lies within a critical volume $v^*$ (of critical radius $R^*$) of the other end of the same molecule, while the relative polar and azimuthal angles of closure ($\theta_N-\theta_0$ and $\phi_N-\phi_0$, respectively) lie below small threshold values ($\theta^*$ and $\phi^*$, respectively). Therefore at short times the concentration of ligated circular monomers will increase with rate
\begin{align}
&\f{\md C(t)}{\md t} = k_0M_0 \notag\\
&\ \ \ \ \ \times P\big(|\biR_N-\biR_0|<R^*, \cos(\theta_N-\theta_0)>\cos\theta^*, \notag\\
& \ \ \ \ \ \ \ \ \ \ \ \ \ \ |\phi_N-\phi_0|<\phi^*\big)\  \ .
\end{align}
Here $k_0$ is the ligation rate for ligation-competent monomers, and $P(|\biR_N-\biR_0|<R^*, \cos(\theta_N-\theta_0)>\cos\theta^*, |\phi_N-\phi_0|<\phi^*)$ is the probability that the chain is considered ligation-competent. The concentration of ligated linear dimers will correspondingly increase as 
\begin{align}
&\f{\md D(t)}{\md t} = 2k_0VM_0^2 \notag\\ 
&\ \ \ \ \ \ \ \ \ \ \times\overline{P}\big(|\biR_N-\biR_0|<R^*, \cos(\theta_N-\theta_0)>\cos\theta^*, \notag\\
&\ \ \ \ \ \ \ \ \ \ \ \ \ \ \ \ \ \ \ \ \ \ \ \ \  |\phi_N-\phi_0|<\phi^*\big) \label{eq:cD} 
\end{align}
where 
\begin{align}
&\overline{P}\big(|\biR_N-\biR_0|<R^*, \cos(\theta_N-\theta_0)>\cos\theta^*, \notag\\
&\ \ \ \ \ \ \ \ \ \ \ \ \ \ \ |\phi_N-\phi_0|<\phi^*\big) = \f{v^*}{V}\left(\f{1-\cos\theta^*}{2} \right)\f{\phi^*}{2\pi}
\end{align}
is the probability of one end of a given molecule lying within the critical volume of the end of another molecule, within the critical relative polar and azimuthal angles. The restriction enzymes used to generate sticky ends  in the experiments discussed here (\emph{Eag}I and \emph{Hind}III) operate on reverse-palindromic sequences, so each monomer has an identical overhanging sequence on each end, and thus either end can be ligated to another monomer, producing the factor of 2 in Eq.~\eqref{eq:cD}.~~\cite{Taylor:1990p37916} 

We assume that end-to-end distances at which ligation occurs are smaller than any length scale characterizing chain statistics, and thus that the probability density is uniform and isotropic within the critical volume and critical angles,
\vspace{0.1em}
\begin{widetext}
\begin{equation}
P(|\biR_N-\biR_0|<R^*, \cos(\theta_N-\theta_0)>\cos\theta^*, |\phi_N-\phi_0|<\phi^*) = v^*(1-\cos\theta^*)\phi^*\ \langle\delta(\biR_N-\biR_0)\, \delta(1-\cos[\theta_N-\theta_0])\, \delta(\phi_N-\phi_0)\rangle_N
\end{equation}
\end{widetext}
where the angled brackets with subscript $N$ denote a canonical average over the $2N$ angular degrees of freedom (two at each internal node) of an $(N+2)$-node chain according to the specified energy function (in our case $E_{\rm chain}$ in Eq.~\eqref{equ:wlce}). The Dirac $\delta$-function $\delta(x)$ is 0 when $x\ne 0$ and integrates to unity over any region including $x=0$, thus the latter two $\delta$-functions vanish unless the tangent vectors $(\biR_1-\biR_0)/d$ and $(\biR_{N+1}-\biR_N)/d$ are identical. (See Fig.~\ref{fig:chain}a.) 

Under these conditions, the $J$-factor reduces to an effective probability density of the unligated chain forming a transient closed loop, with zero relative polar and azimuthal angles:
\begin{subequations}
\begin{align}
J &= 2\f{k_{\rm cyc}}{k_{\rm dim}}\\
&= 2M_0\, \lim_{t\rightarrow 0} \left[ \f{\md C(t)}{\md t} \Big/ \f{\md D(t)}{\md t}\right]\\
&= \f{4\pi}{v^*(1-\cos\theta^*)\phi^*}\notag\\
&\ \ \ \ \ \times  P\big(|\biR_N-\biR_0|<R^*, \cos(\theta_N-\theta_0)>\cos\theta^*, \notag\\
&\ \ \ \ \ \ \ \ \ \ \ \ \ \ \ \ \ \ \ \ |\phi_N-\phi_0|<\phi^*\big) \\
&= 4\pi\ \langle\delta(\biR_N-\biR_0)\, \delta(1-\cos[\theta_N-\theta_0])\, \delta(\phi_N-\phi_0)\rangle_N\ . \label{eq:jCK}
\end{align}
\end{subequations}
In \ts\ref{sec:sims} we compute the probability density $\langle\delta(\biR_N-\biR_0)\, \delta(1-\cos[\theta_N-\theta_0]) \, \delta(\phi_N-\phi_0)\rangle_N$ for the meltable WLC model outlined in \ts\ref{sec:modelMWLC}. 

When ligase concentration is not sufficiently low, the ligation rate of ligation-competent conformations becomes comparable to the rate of their formation from ligation-incompetent precursors. In this concentration regime ligation rates will thus depend on the dynamics of formation of ligation-competent conformations, not just on their equilibrium probabilities. For circular monomers, this reflects the rate of motion in the coordinate of the end-to-end distance. For linear dimers, kinetics will be diffusion-controlled. Thus the ratio of circularization and dimerization rates no longer simply reflects an equilibrium probability, and the $J$-factor no longer reflects equilibrium free energies. Du, \emph{et al.} {\add showed that at 21\deg C the threshold ligase concentration was $\sim$100 units/ml, and thus argued that Cloutier and Widom's ligase concentrations of 150-250 units/ml were sufficiently high that they did not measure equilibrium bending propensities. Recent experiments by Forties, et al.~\cite{Forties:2009p61935} complicate the picture by indicating a threshold at 37\deg C of ~$\sim$400 units/ml, suggesting that Cloutier and Widom's experiments at 30\deg C might have near-threshold ligase concentrations.  See also Peters and Maher's review for further discussion.~\cite{Peters:2010p60601} }

In the {\add limit} of high ligase concentration, the rate-limiting step is formation of a ligation-competent conformation, and hence the $J$-factor would reflect the effective reaction rate of aligning the ends. This quantity can in principle be calculated from molecular dynamics simulations of models with more detailed representation of the DNA: though the relevant dynamical modes might be complex, a first approximation might assume diffusive motion in the end-to-end distance and join angle, and subsequently solve for the diffusion-controlled steady state using a Langevin or Fokker-Planck approach.  Comparison with experiments at high ligase concentration could thus yield insights into DNA dynamics on these short length scales.  Cheng, \emph{et al.} recently conducted similar experimental and computational examination of the cyclization dynamics of single-stranded DNA.~\cite{Cheng:2010p65996} {\add In the rest of this paper, we restrict our discussion to the regime of low ligase concentration.}

\section{Connection with lattice models \label{sec:graft}}
One-dimensional lattice models for fluctuations in DNA hybridization have a long history in the study of bulk melting thermodynamics.~\cite{braggZimm} Systems of interest are typically free of any external mechanical constraints, so the models typically make no statement about chain structure. In particular they assign free energy to a short segment of the chain based only on its hybridization state $m$:
\be
\overline{E}(m) = \delta_{m,1} \, \ODm(T).
\label{equ:lm}
\ee
The chain's total free energy in this description similarly depends only on the hybridization variables, $\overline{E}_{\rm chain}[\{m_i\}] = \sum_i \overline{E}(m_i)$. Implicit in the free energy assignment of Eq.~\eqref{equ:lm} is an integration over conformational fluctuations weighted by the Boltzmann distribution,
\begin{align}
&\exp\!\left(-\beta \overline{E}_{\rm chain}[\{n_i\}]\right) \\
& \ \ \ \ \ = \int_{-1}^1 \!\!\!\! d(\cos\theta_1) \int_{-1}^1 \!\!\!\! d(\cos\theta_2) \, \ldots \int_{-1}^1 \!\!\!\! d(\cos\theta_{N-1}) \notag\\
& \ \ \ \ \ \ \ \ \ \ \ \ \ \ \ \ \ \ \ \ \times\exp\!\left(-\beta E_{\rm chain}[\{n_i\}, \{\theta_i \}]\right). \notag
\end{align}
Analytical evaluation of this integral is straightforward for the meltable WLC, since the Boltzmann weight factorizes. 

The partition function at a given node for hybridization state $m$ (integrating Eq.~\eqref{equ:mwlc} over all bend angles $\cos\theta$) is
\begin{subequations}
\begin{align}
q(m) &= \int_{-1}^1 \md(\cos\theta) \, e^{-\beta E(m,\cos\theta)}\\
&= \f{d}{\lPH + \delta_{m,1}\left(\lPM-\lPH\right)} \exp\left\{-\delta_{m,1}\bDm\right\} \\
& \qquad \times \left(1 - \exp\left\{-\f{2}{d}\left[\lPH + \delta_{m,1}\left(\lPM-\lPH\right)\right]\right\}\right). \notag
\end{align}
\end{subequations}
The free energy difference between melted ($m=1$) and non-melted ($m=0$) states is $\Delta F = F(1) - F(0) = -\kT\, \ln[q(1)/q(0)].$ We arrive at a relationship between the `bare' free energy of melting ($\Dm^{\rm {\add MWLC}}$) and its `renormalized' counterpart ($\ODm$) that accounts for the influence of bending fluctuations:
\begin{subequations}
\label{equ:renorm}
\begin{align}
\label{equ:renormExact}
\bODm &= \bDm^{\rm {\add MWLC}}  - \ln\!\left[\f{\lPH}{\lPM} \left(\f{1 - \exp(-2\lPM/d)}{1 - \exp(-2\lPH/d)}\right)\right] \\
&\approx \bDm^{\rm {\add MWLC}} - \ln\left(\f{\lPH}{\lPM}\right).
\end{align}
\end{subequations}
The second term in {\add the RHS of Eq.~\eqref{equ:renormExact}} accounts for the greater range of bending motion available to a molten segment of the chain, which is considerable. For the length parameters employed by Yan and Marko~\cite{Yan:2004p1403} ($\lPH=50$ nm, $\lPM=1$ nm, $d=1$ nm), the renormalization evaluates to
\be
\ODm - \Dm^{\rm {\add MWLC}} \approx -3.8\,\kT\ .
\label{equ:rval}
\ee
{\add Similar analysis for the KWLC yields
\begin{align}
\bODm &= \bDm^{\rm KWLC} - \ln\left(\f{2\, \lPH}{d[1-\exp(-2\, \lPH/d)]}\right) \\
&\approx \beta\Delta\mu^{\rm KWLC} - \ln \left( \f{2\, \lPH}{d} \right) \\
&\approx \beta\Delta\mu^{\rm KWLC} - 4.6 \ ,
\end{align}
and for the spontaneous bend model
\begin{align}
\bODm &= \bDm^{\rm SB} - \ln\sqrt{\f{\pi}{2\, d\,  \lPM}} \f{\lPH}{[1-\exp(-2\, \lPH/d)]} \\
&\times \left\{ \text{erf}\left[ \sqrt{\f{\lPM}{2d}}(\gamma+1) \right] - \text{erf}\left[ \sqrt{\f{\lPM}{2d}}(\gamma-1) \right] \right\} \ .
\end{align}
}
Conversely, integrating over melt degrees of freedom renormalizes the persistence length.~\cite{Palmeri:2007p50691,Mastroianni:2009p46259}

\section{Parameterization \label{sec:param}} 
The large body of work on DNA melting provides good estimates for the thermodynamic parameters appropriate to one-dimensional lattice models such as Eq.~\eqref{equ:lm}. {\add Bulk melting free energies include a renormalization [Eq.~\eqref{equ:renorm}] due to integrating out bending fluctuations, therefore proper use of melting data to parameterize microscopic models must involve disentangling this renormalization.} If each node or lattice site represents three basepairs, then we expect $\ODm\approx 8-11\, \kT$ at ambient conditions.~\cite{SantaLuciaJr:1998} Yan and Marko presented an argument for this range of values,~\cite{Yan:2004p1403} but no distinction was made there between $\Dm^{\rm {\add MWLC}}$ and $\ODm$. If we assign the inferred melting penalty to the bare free energy rather than the renormalized value (\emph{i.e.}, $\Dm^{\rm {\add MWLC}}\approx 8-11\, \kT$ and $\ODm\approx 4-7\, \kT$), hybridization statistics of the meltable WLC at room temperature are not consistent with experiment. Specifically, a three-base region is melted with a probability between $7.2\times 10^{-4}$ and $1.4\times 10^{-2}$ (\emph{i.e.}, between $e^{-4.2}$ and $e^{-7.2}$), instead of between $1.7\times 10^{-5}$ and $3.3\times 10^{-4}$ (\emph{i.e.}, between $e^{-8}$ and $e^{-11}$) as suggested by thermodynamic measurements.

Furthermore, this significantly-increased probability of melts yields chains that are too compliant to be consistent with longer length scale force-extension experiments. The persistence length is determined by $\avCos$, the average cosine of the angle at a given node. For a meltable node this average cosine is an equilibrium-weighted linear combination of the average cosines of a melted or hybridized node, 
\be
\avCos = \f{e^{-\bODm}}{1+e^{-\bODm}}\avCos_{\rm M} + \f{1}{1+e^{-\bODm}}\avCos_{\rm H}\ ,
\ee
producing {\add for the MWLC} an apparent persistence length (averaged over melt fluctuations)
\be
\lP^{\rm {\add MWLC}} = \f{d}{1-\avCos^{\rm {\add MWLC}}} = \f{1+e^{-\bODm}}{(e^{-\bODm}/\lPM) + (1/\lPH)}\ .
\ee
In particular, for the range of bare melt free energies $\Dm^{\rm {\add MWLC}}= 8-11\, \kT$, the persistence length is renormalized to $\lP^{\rm {\add MWLC}}\approx$ 30-48 nm.  

To maintain a given apparent persistence length $\lP$ for a given renormalized melt free energy $\bODm$, the hybridized persistence length $\lPH$ is thus a function of the melt persistence length $\lPM$: 
\be
\lP^{\rm H,{\add MWLC}} = \f{\lP}{1 - \left(\f{\lP}{\lPM}-1\right)e^{-\bODm}}\ .
\label{equ:lPHbODm}
\ee

{\add For the KWLC model we find
\be
\lP^{\rm H, {\add KWLC}} = \f{\lP}{1 - \left(\f{\lP}{d}-1\right)e^{-\bODm}}\ ,
\ee
and for the spontaneous bend model with at least modest melt persistence length
\be
\lP^{\rm H, {\add SB}} \approx \f{\lP}{1 - \left(\f{\lP}{d}[\gamma-1]-1\right)e^{-\bODm}}\ .
\ee
}

Based on Eqs.~\eqref{equ:rval} and~\eqref{equ:lPHbODm} we determine a more realistic value for the bare free energy of local melting {\add in the MWLC}, $\Dm^{\rm {\add MWLC}} \approx 12-15\, \kT$, which is consistent with bulk thermodynamics by construction. The increased $\Dm^{\rm {\add MWLC}}$, designed to avoid double-counting entropic gains due to increased flexibility, will clearly act to suppress melting and to diminish its importance in the kinetics of cyclization. We have quantified this suppression using computer simulations.

\section{Relating $J$-factors to distributions of end-to-end distance and angle in a discretized chain \label{sec:relatingJ}}
The $J$-factor was expressed in \ts\ref{sec:cycKin} in terms of an equilibrium probability density that lends itself to straightforward evaluation in computer simulations. For a chain of contour length $(N+1)d$, whose $N+2$ nodes are located at positions $\{\biR_0,\biR_1,\biR_2,\ldots,\biR_{N+1}\}$, we rewrite Eq.~\eqref{eq:jCK} as 
\be
J(N) = 4\pi\f{Q_N^{({\rm loop})}}{Q_N^{(0)}}\ ,
\label{equ:jprob}
\ee
We have defined partition functions $Q_N^{({\rm loop})}$ for a cyclized $(N+2)$-node meltable WLC, and $Q_N^{(0)}=q^N$ for the same molecule absent end constraints. The single-node partition function $q$ can be evaluated as
\be
q = 2\pi \int_{-1}^1 \!\!\! d(\cos\theta) \,\, e^{-\beta \tilde{E}(\cos\theta)},
\ee
for {\add apparent} bending energy
\be
\tilde{E}(\cos\theta) = -\kT\, \ln \sum_{n=0}^1 e^{-\beta E(m,\cos\theta)}.
\ee
Thus,
\begin{align}
&J(N) = 4\pi \int\md(\cos\theta_1)\cdots\int\md(\cos\theta_N)\\ 
&\ \ \ \ \ \ \ \ \ \ \ \times\int\md\phi_1\cdots\int\md\phi_N \f{e^{-\beta\sum_{j=1}^N\tilde{E}(\cos\theta_j)}}{q^N} \notag\\
&\ \ \ \ \ \ \ \ \ \ \ \times\delta(\biR_N-\biR_0) \, \delta(1-\cos[\theta_N-\theta_0]) \, \delta(\phi_N-\phi_0) \notag
\end{align}
The first $\delta$-function imposes no constraint on the variables $\cos\theta_N$ or $\phi_N$ (see Fig.~\ref{fig:chain}a), and thus integrating over both is trivial, giving
\be
J(N) = 4\pi q^{-1} \, \left\langle\delta(\biR_N-\biR_0)\, e^{-\beta\tilde{E}(\cos\theta_0)} \right\rangle_{N-1}\ .
\ee
We then use the delta-function identity 
\begin{align}
&e^{-\beta\tilde{E}(\cos\theta_0)} = \\
&\ \ \ \ \  \int\md(\cosThetaJoin) \,e^{-\beta\tilde{E}(\cosThetaJoin)\, \delta(\cosThetaJoin-\cos\theta_0)}\notag\ ,
\end{align}
where the join angle $\thetaJoin$ describes the relative orientations of the chain's ends, $\cosThetaJoin = (\biR_N-\biR_{N-1}) \dot (\biR_1-\biR_0) / d^2$, when it is poised for cyclization, $\biR_N = \biR_0$ (i.e., $\biR = {\bf 0}$). (See Fig.~\ref{fig:chain}b.) Finally, we multiply and divide by $\langle\delta(\biR_N-\biR_0)\rangle_{N-1}$, substitute the probability density at zero end-to-end extension, $p_N(\biR_0=\biR_N) = \langle\delta(\biR_N-\biR_0)\rangle_{N-1}$, and introduce the conditional probability density at a given bend angle of a particular node (given a cyclized chain), 
\begin{align}
\lefteqn{ p_N(\cosThetaJoin\, |\, \biR={\bf 0}) =} \\
&\ \ \ \ \ \ \ \ \ \ \ \ \ \ \ \langle\delta(\biR_N-\biR_0) \, \delta(\cosThetaJoin - \cos\theta_0)\rangle_{N-1} \notag\\
&\ \ \ \ \  \ \ \ \ \ \ \ \ \ \ \ \ \ \ \ \ \ \ \ \ \ \ \ \ \ / \  \langle\delta(\biR_N-\biR_0)\rangle_{N-1}\notag\ .
\end{align}
These manipulations enable us to express the $J$-factor in terms of quantities readily estimated from sampling of canonically distributed microstates,
\begin{align}
J(N) &= 4\pi q^{-1}\ p_N(\biR={\bf 0}) \int\md(\cosThetaJoin) \label{equ:jcalc}\\
&\ \ \ \ \ \ \ \ \ \ \times e^{-\beta\tilde{E}(\cosThetaJoin)} \, p_N(\cosThetaJoin\, |\, \biR={\bf 0})\notag\ .
\end{align}

\section{Monte Carlo simulations \label{sec:sims}}
Eq.~\eqref{equ:jprob}, together with an expansion of the Boltzmann distribution in spherical harmonics, was used by Yan and Marko \cite{Yan:2004p1403} to evaluate $J(N)$ for the meltable WLC. Their approach becomes quite numerically demanding, however, for the short chains of interest here. We have instead used Monte Carlo simulations to estimate the probability distributions appearing in Eq.~\eqref{equ:jcalc}. This approach also permits the simple extension of the model to more complicated energetics, involving considerations such as volume exclusion and DNA helicity.~\cite{Mastroianni:2009p46259}

We employ two kinds of trial moves to efficiently navigate conformational space while maintaining the geometric constraints of the molecule. `Free' rotations divide the chain in two by selecting a randomly chosen node, then rotate the two segments relative to one another about a randomly chosen axis through the dividing node. ÔCrankshaftÕ moves rotate a segment of the chain between two randomly chosen nodes, about the axis connecting them. The number of contiguous nodes rotated in crankshaft moves is chosen from a uniform distribution on the interval [1,10]. Angles for all rotations are selected at random from Gaussian distributions whose variances were adjusted until roughly $1/2$ of all trial moves are accepted. Melt moves attempt to change the hybridization state of a randomly chosen basepair, either from hybridized to melted or \emph{vice versa}. 

Calculating cyclization probabilities for short DNA chains via standard equilibrium sampling is computationally challenging because of the extreme energetic cost of juxtaposing the two ends with the correct orientation. We estimated the equilibrium average on the right-hand side of Eq.~\eqref{equ:jprob} by sampling from the Boltzmann distribution using Wang-Landau sampling,~\cite{Landau:2004p8092} an adaptive method for broadly sampling fluctuations of a chosen order parameter without any prior knowledge about the details or even the gross shape of the probability distribution. 

\begin{figure*}[!htbp]
\centering
\includegraphics{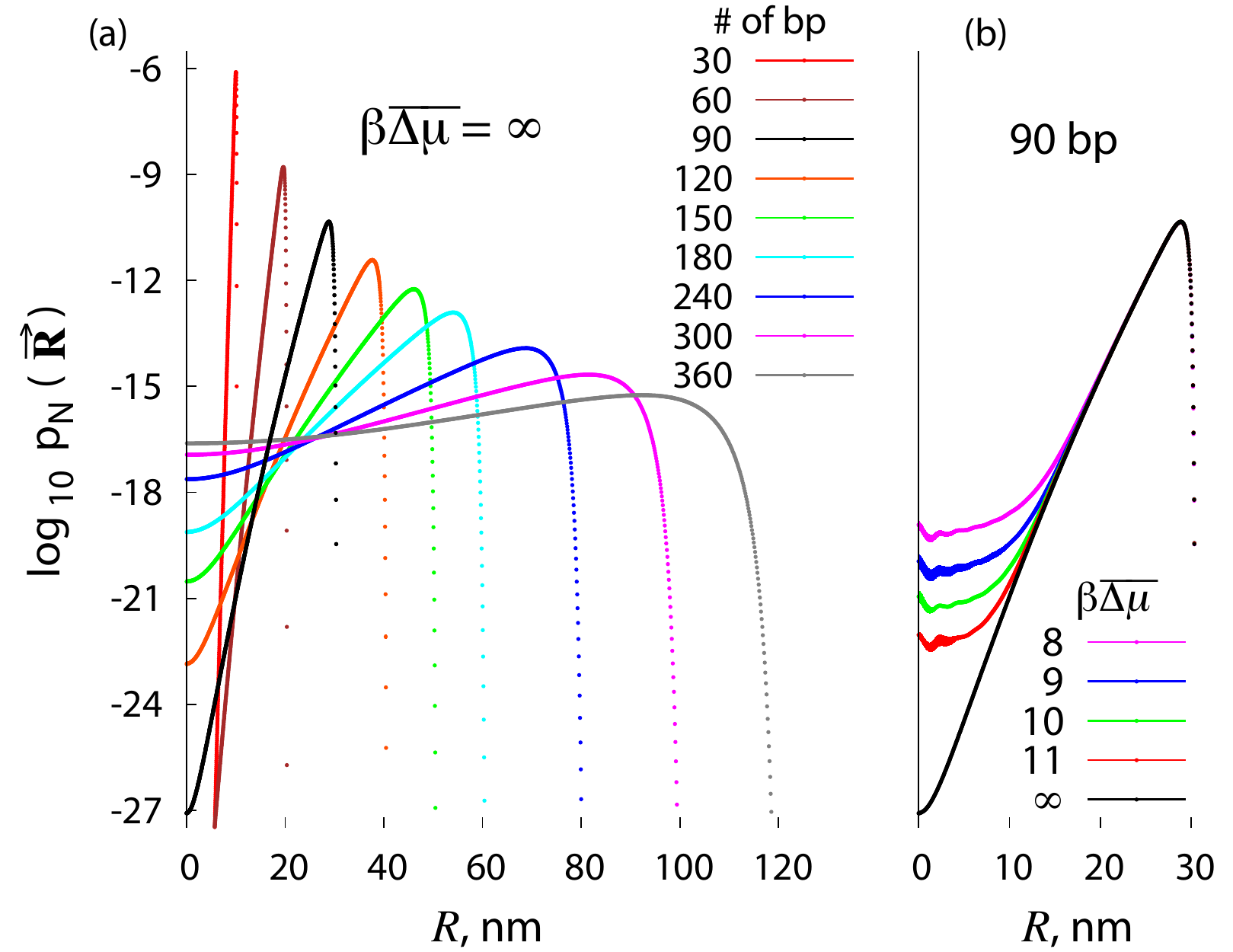}
\caption{Bending statistics for meltable wormlike chains of varying contour length and melting thermodynamics. Error bars show $\pm$ 2 standard errors of the mean (s.e.m.), calculated from three independent simulations.  (a) Distributions of end-to-end distance $R$, for chains comprising $30-360$ bp, generated by Wang-Landau sampling. Melts are disallowed here, i.e., $\bODm = \infty$.  (b) Distributions of end-to-end distance $R$, for 90 bp meltable wormlike chains for several values of the (renormalized) free energy of melting, $\ODm$.} \label{fig:jFSpace}
\end{figure*}

\begin{figure*}[!htbp]
\centering
\includegraphics{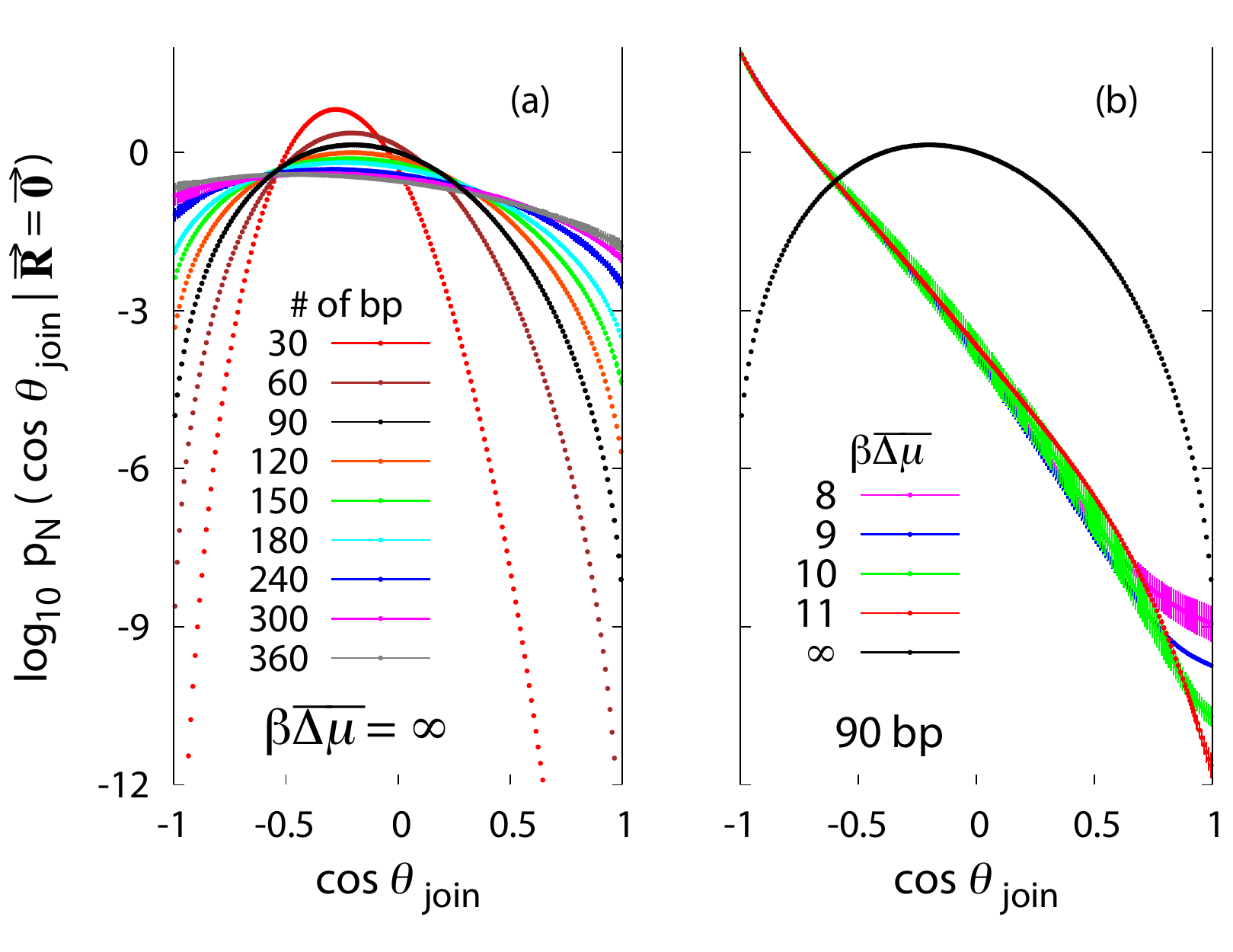}
\caption{Statistics of chain's initial orientation ($\theta_0$), relative to that of its final link ($\theta_N$), for cyclized meltable wormlike chains of varying contour length and melting thermodynamics. Error bars show $\pm$ 2 s.e.m.  (a) Distributions of $\cosThetaJoin$, for wormlike chains comprising $30-360$ bp, generated by Wang-Landau sampling. Melts are disallowed here, i.e., $\bODm = \infty$.  (b) Distributions of $\cosThetaJoin$ for 90 bp meltable wormlike chains for several values of the (renormalized) free energy of melting, $\ODm$.} \label{fig:jFAng}
\end{figure*}

This biased sampling approach allowed us to construct the distribution $p_N(\biR)=\langle \delta(\biR-(\biR_N-\biR_0)) \rangle_N$ of the end-to-end vector $\bf{R}$ for an uncyclized chain, over the full range of $|\biR|$ (Fig.~\ref{fig:jFSpace}). Similarly biased sampling, this time of a cyclized chain with no bending energy at one node, was used to construct the conditional angular distribution at that freely-jointed node (Fig.~\ref{fig:jFAng}):
\vspace{0.1em}
\begin{widetext}
\begin{equation}
p_N( \cosThetaJoin\ |\ \biR={\bf 0}) = \f{  \left\langle \delta \left(\cosThetaJoin - d^{-2}[\biR_1-\biR_0]\cdot[\biR_N-\biR_{N-1}]\right) \: \delta(\biR_N-\biR_0) \right\rangle_N  }{  \left\langle\delta(\biR_N-\biR_0)\right\rangle_N  } \ .
\end{equation}
\end{widetext}
For each chain length and melting free energy, 3 different sampling trajectories were run from identical initial conditions but with different random number sequences to calculate uncertainties. Each Wang-Landau iteration ran until each of the 1000 spatial bins (or 200 angular bins) was visited at least 100 times. The 10 spatial bins at greatest extension were excluded from this criterion due to their exceedingly small probabilities. The increment to the Wang-Landau bias began at $1\, \kT$ and decreased by 10\% at each subsequent iteration, until the increment reached $10^{-7}\, \kT$ ($10^{-8}\, \kT$ for angular simulations of chains shorter than 120 bp). A few angular simulations of shorter chains achieved greater precision with a 20\% (rather than 10\%) decrease in increment per iteration. 

Probability distributions of the end-to-end distance $\biR$ and join angle $\theta_{\rm join}$, for meltable wormlike chains of varying length and melt free energy, are shown in Figs.~\ref{fig:jFSpace} and \ref{fig:jFAng}. For meltable chains, the end-to-end distribution approximates at large distances that of the un-meltable chain. As end-to-end distance decreases below the range of typical fluctuations of a WLC, melting begins to significantly facilitate achieving loop closure of sufficiently short chains. Their significant energetic cost requires that substantial bending forces develop before such local excitations become favorable. 

The general shape of $p_N(\biR)$, which has been discussed at length elsewhere,~\cite{Allemand:2006p7102} can be simply understood in terms of entropic elasticity at large $\biR$ and bending energy at small $\biR$. Straightening a chain imposes high entropic cost, so that probability decreases sharply with $\biR$ at large extension. Imposing a short end-to-end distance requires significant bending and thus high enthalpic cost, explaining the low probability for very bent chains. The longest chain is easiest to cyclize since the total bend can be distributed over the largest number of basepairs.  

At the very shortest end-to-end distances, meltable chains show a slight increase in $p_N(\biR)$ as distance decreases, for geometric reasons. Imposing such a tight bend essentially requires a melt at the middle node of the chain; once that node is melted it serves as a nearly free joint between two nearly rigid chain segments on either side. Taking the joint as the origin of a molecular reference frame, the two endpoints thus effectively explore the surface of a sphere. As they approach one another in this two-dimensional geometry, the number of accessible conformations $\Omega$ with a given value of $R$ grows only linearly, $\Omega \sim R$ (rather than quadratically, as it would for a very flexible chain). The probability per unit volume $p_N(\biR)$ thus scales here roughly as $\Omega(R) / (4\pi R^2) \sim 1/R$. Because the chain ends are not strictly confined to a two-dimensional surface, $p_N(\biR)$ does not diverge as $R \rightarrow 0$, but it does grow over a small range of $R$. A similar scenario, now involving melts at nodes one or two removed from the central one, accounts for the small oscillations at slightly longer end separations.  Consistent with this explanation, the local maxima in probability appear at end-to-end distances that are even integer multiples of the link length.  

For unmeltable chains, the join angle statistics (for a chain that meets at its ends but lacks bending energy at that meeting point) in Fig.~\ref{fig:jFAng} show a probability maximum at intermediate angles.  This stems from the enthalpic preference for equally distributing the bend over the entire contour of the chain, producing a teardrop-shaped chain contour with join angle intermediate between 0\deg (straight) and 180\deg (fully bent).~\cite{Yamakawa:1972p60609} Meltable short chains differ substantially. In accommodating loop closure, the energetic cost of severe bending distributed across the entire uniformly curved chain (without any melts) is substantially higher than that of a single melt at the middle basepair opposite the junction. The latter state is energetically minimized with a join angle of $\sim$180\deg\, and very little bending on either side. Conversely, the low probability of a join angle of 0\deg ($\cosThetaJoin = 1$) reflects the need for formation of a second melt, essentially required to bring the ends of such short chains into common orientation.

\section{$J$-factors \label{sec:jF}}
We have confirmed numerically that this MC approach is consistent with the spherical harmonic expansion used by Yan and Marko.~\cite{Yan:2004p1403} For the chains we compared, of length 135-225 bp, the deviations between their results and our calculations are not statistically significant (data not shown). Henceforth we calculate {\add MWLC} $J$-factors for parameters that differ in two respects from those of Yan and Marko~\cite{Yan:2004p1403}: (i) melted sections are slightly stiffer, $\lPM = 2.5$ nm, than in their calculations (where $\lPM = 1$ nm); and (ii) the range of melting free energy values we explore extends to the stiffer thermodynamic penalties suggested in \ts\ref{sec:param}. We argue that the longer persistence length of unhybridized chains is more consistent with experimental estimates for single-stranded DNA (ssDNA) bending elasticity,~\cite{Smith:1996p21972,Murphy:2004p7356} since a melted section composed of two single strands of DNA in close apposition would be expected to have a persistence length at least twice that of ssDNA. 

All of the model results we report, like those of Yan and Marko, neglect the helical nature of hybridized dsDNA, implicitly assuming that two ends juxtaposed with any helical phasing are ligation-competent. The calculations thus represent an upper bound for the $J$-factor of a helical wormlike chain polymer, a bound that is tightest for chains that are an integral number of helical repeats in length. 

\begin{figure*}[!htbp]
\centering
\includegraphics{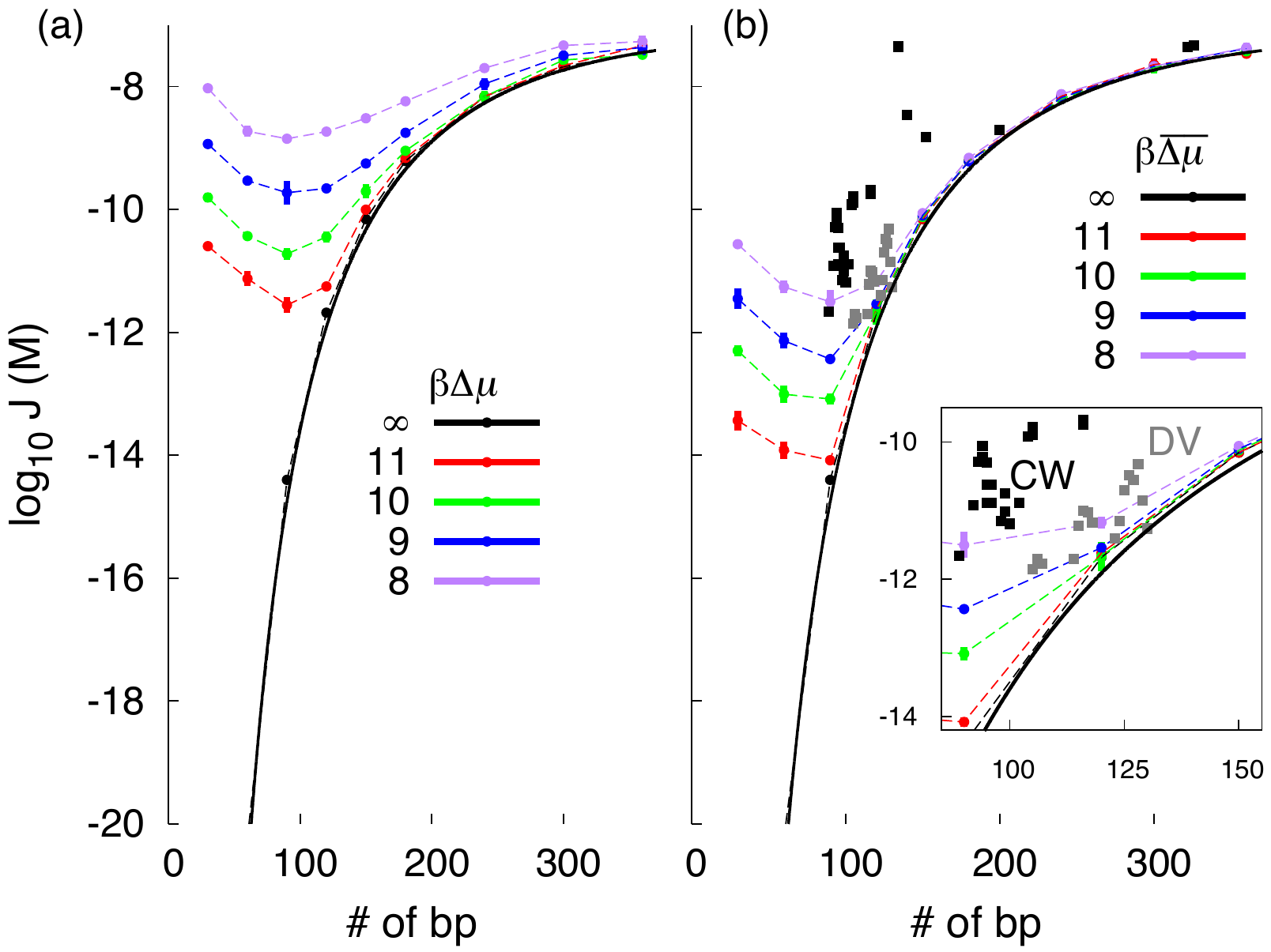}
\caption{Cyclization kinetics calculated for meltable wormlike chains of varying contour lengths and melting thermodynamics. Error bars show $\pm$ 2 s.e.m. (a) $J$-factor plotted as a function of contour length, for {\em bare} melting free energies $\Dm = 8\, \kT$ (magenta), $9\, \kT$ (blue), $10\, \kT$ (green), $11\, \kT$ (red), and in the absence of melts, $\Dm = \infty$ (black). (b) $J$-factor plotted as a function of contour length, with {\em renormalized} melting free energies $\ODm = 8\, \kT$ (magenta), $9\, \kT$ (blue), $10\, \kT$ (green), $11\, \kT$ (red), and in the absence of melts, $\ODm=\infty$ (black). Solid black line shows predictions from the approximate analytic theory of Shimada and Yamakawa (in the absence of melts).~\cite{Shimada:1984p37932} Black squares show experimental results of Cloutier and Widom;~\cite{Cloutier:2004p4086} gray squares those of Du, \emph{et al.}~\cite{Du:2005p1748}
\label{fig:jBareRenorm}}
\end{figure*}

Fig.~\ref{fig:jBareRenorm} shows our numerical results for contour lengths of 10-120 nm, with melting energies from Yan and Marko~\cite{Yan:2004p1403} (a) or with values of $\Dm$ that are consistent with bulk melting thermodynamics (b). The shift from $\Dm = 8-11\, \kT$ (Fig.~\ref{fig:jBareRenorm}a) to $\ODm = 8-11\, \kT$ (Fig.~\ref{fig:jBareRenorm}b) represents an increase in melting free energy of $3.7\, \kT$ and hence considerably suppresses melts. In fact, whereas for $\Dm = 8-11\, \kT$ only the 360 bp chain $J$-factors are unaffected by melts, for $\ODm = 8-11\, \kT$ ($\Dm = 11.7-14.7\, \kT$) meltable chains as short as 150 bp cyclize with propensities indistinguishable from the melt-free limit, $\ODm = \Dm = \infty$. For shorter chains the meltable models produce a $J$-factor significantly greater than that for the melt-free ($\Dm=\ODm=\infty$) model, because cyclization dictates a bend so sharp that the energetic gain from localizing the bend at a floppy melted section overcomes the energetic cost of forming the melt. 

Surprisingly, at the short chain lengths (less than $\sim$90 bp), a meltable chain becomes easier to cyclize with shorter chain length, reflected in a higher $J$-factor. This trend is rationalized through consideration of the enthalpic and entropic costs of cyclization. Cyclizing a short fully-hybridized DNA is so energetically costly (seen in the sharp drop in $J$-factor for the un-meltable chain (black curve)) that a short meltable chain essentially requires a melt near the middle to cyclize. Such a cyclized structure with a single melt will have an enthalpy largely independent of chain length, due to the high compliance of the melted region. However the entropic cost of cyclization decreases with decreasing chain length, so for short meltable chains the probability of cyclization increases with decreasing chain length. 

{\add This striking increase in $J$-factor for very short chains also occurs for the KWLC.~\cite{Wiggins:2005p1466} It does not appear in the spontaneous bend model with sufficiently rigid molten sections, nor in a model that fixes a particular bend angle~\cite{Du:2005p1748} (equivalent to the spontaneous bend model with infinite molten persistence length $\lPM$), because in these models cyclization of shorter chains still requires tighter bending and thus greater enthalpic cost. It also would not occur for the linear subelastic chain~\cite{Wiggins:2006p1461} as that model does not concentrate bending at one location. No experiments have yet probed the short lengths where this phenomenon is seen for the MWLC or KWLC.}

We now compare with experiment these computational results for meltable wormlike chains with the stiffer renormalized melting free energies $\ODm = 8-11\, \kT$ (Fig.~\ref{fig:jBareRenorm}b inset). Cloutier and Widom's experimental cyclization rates \cite{Cloutier:2004p4086} show significant scatter, presumably reflecting sequence heterogeneity in melt- and bend- propensities even among `random' sequences (which exclude nucleosome positioning sequences known to have unusual bending characteristics). The meltable wormlike chain model predicts cyclization rates significantly lower than those inferred from experiment over the range of 90-120 bp, which includes the majority of Cloutier and Widom's sequences. Melting excitations, when made thermodynamically consistent with bulk melting behavior, appear insufficient to explain their results. 

The experimental results of Du, \emph{et al.}~\cite{Du:2005p1748} highlight the $J$-factor's strong helical dependence at these short lengths.  Local maxima in $J$ occur for chain lengths that are integer multiples of the helical repeat,  thus requiring minimal over- or under-twisting to bring apposing ends into helical register. These chain lengths should offer the most straightforward comparison with the meltable wormlike chain model (which implicitly assumes perfect helical register). Throughout the range of $N$ studied in these experiments, these maxima are slightly higher than corresponding $J$-factors of the meltable wormlike chain, or compare favorably with results for the smallest values of $\Delta\mu$. Given debate over the precise value for DNA persistence length within the range of $\sim$45-53 nm, the experiments of Du, \emph{et al.} appear to be consistent with a meltable wormlike chain model. But in this range of uncertainty, their results are almost equally consistent with an unmeltable wormlike chain, which has similar $J$-factors to the meltable wormlike chain for molecules longer than $\sim$120 bp. Shorter chains would offer a much more discriminating test: for $N \ltsim$ 50 bp, the predicted $J$-factors of meltable and un-meltable chains differ enormously. Experiments on such molecules should unambiguously address the importance of thermally-excited melts in the severe bending of dsDNA.

\section{Temperature dependence}
Enhanced cyclization of long DNA strands (thousands of bp) due to elevated temperature was established decades ago.~\cite{Mertz:1972p32854} {\add Yet there are few experiments (to our knowledge) probing temperature-variation of cyclization kinetics at the short lengths of interest here.} Given the expected strong Arrhenius temperature-dependence of melting, such measurements {\add should} clarify the role of dehybridization or even explain discrepancies between different experiments. Here we scrutinize these possibilities through the thermal sensitivity of an MWLC. 

In this section we present numerical results for $J$-factors as a function of $T$. We assume (i) temperature-independent bending rigidities $\kT\,\lPH$ and $\kT\,\lPM$; and (ii) a constant value of the apparent bending rigidity $\kT\,\lP$, a weighted combination of hybridized and melted bending rigidities that is dominated by the former. Transient electric birefringence experiments in the range of 20-43\deg C~\cite{Lu:2001p7077} support this latter assumption. 

\begin{figure}[!htbp]
\centering
\includegraphics{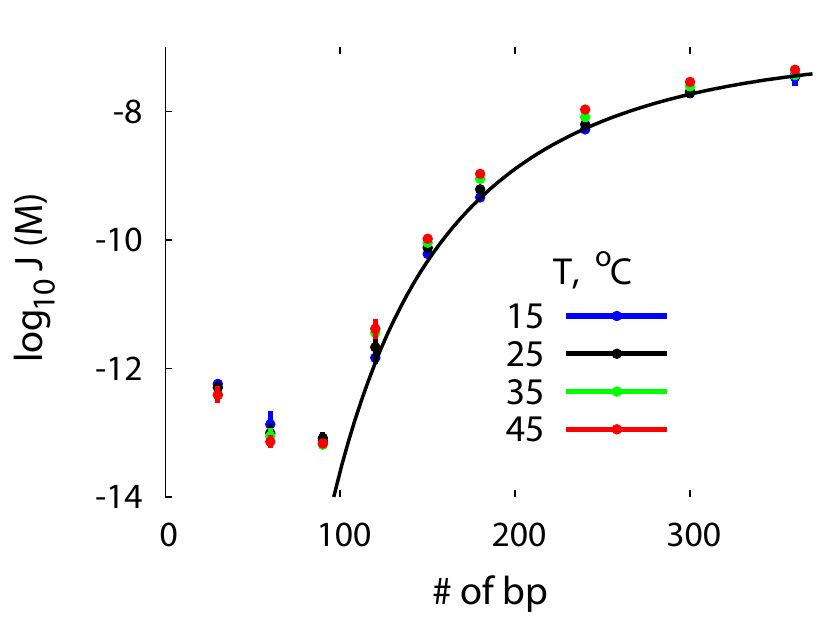}
\caption{Temperature dependence of cyclization kinetics for meltable wormlike chains. $J$-factors are plotted as functions of contour length for $T$ = 15, 25, 35, and 45\deg C. A moderate melt enthalpy, $\ODm = 10\, \kT$, was used in all cases. Error bars show $\pm$ 2 s.e.m. Solid black line shows predictions from the approximate analytic theory of Shimada and Yamakawa (in the absence of melts).~\cite{Shimada:1984p37932} \label{fig:jTDep}}
\end{figure}

Fig.~\ref{fig:jTDep} shows computed $J$-factors, for the case of $\ODm = 10 \kT$, at several different temperatures. We find that a change of 10\deg C has very little impact on $J$ for all chain lengths examined. In fact varying temperature by as much as 30\deg C never altered $J$ by more than a factor of three, for all lengths examined. {\add Experiments spanning 23-42\deg C found similar temperature-sensitivity for a 200 bp segment of $\lambda$ DNA and a 116 bp chain with sequence chosen to minimize melting, but somewhat greater temperature-sensitivity in a 116 bp chain with sequence designed to accentuate melting.~\cite{Forties:2009p61935}} Since the temperatures in the respective experiments of Du \emph{et al.} and Cloutier and Widom differ by only {\add nine degrees} C, temperature variation is unlikely to explain differences in their findings.

\section{Conclusion}
{\add In this work we have explored the mutual influence of distinct degrees of freedom characterizing the microscopic state of a chain molecule. Our results highlight how the presence or absence of one fluctuating variable shapes the effective forces on another. From a practical perspective, these results caution that grafting an additional degree of freedom onto a model changes the apparent parameters governing the fluctuations of other degrees of freedom. 

These concepts find a concrete example in models of DNA flexibility that incorporate regions of thermally-enhanced flexibility.} It seems immediately clear that introducing the possibility of flexible {\add excitations} to a polymer will reduce its {\add apparent} persistence length, hence the invocation of melts in attempts to understand anomalously high cyclization rates. Somewhat more subtly, adding bending degrees of freedom (that differ in their natural fluctuations between hybridized and melted sections of the chain) to each site in a lattice model for melting can substantially renormalize the free energy of dehybridization. For the particular parameters used in our simulations ($\lPH=50$ nm, $\lPM=2.5$ nm, $d=1$ nm), the bending-induced increase in {\add apparent} melt free energy amounts to $3.7\, \kT$. 

{\add Based on this distinction between bare and apparent melt free energies,}  we develop a thermodynamically consistent treatment of the meltable wormlike chain model that properly accounts for the greater entropy of more flexible melted sections, and thus permits accurate parameterization from melting experiments. For sufficiently short DNA chains, the conformations that lead to cyclization are exceedingly rare and thus their equilibrium probability is greatly facilitated by the transiently enhanced flexibility of thermally-excited melts, even at the energetic cost of disrupting DNA basepairing and stacking interactions. However, the proper parameterization of melting energies in the range $\ODm = 8-11\, \kT$ pushes the onset of such melt-enhanced cyclization to significantly shorter lengths, so much so that the cyclization propensity of a meltable chain is practically indistinguishable from an un-meltable chain for chain lengths greater than $\sim$120 bp. Predictions for meltable and un-meltable chains diverge significantly around $\sim$80 bp, suggesting that experiments on even shorter chains would distinguish between models. Indeed at the shortest lengths, meltable chains even show an \emph{increase} in cyclization propensity with decreasing chain length, due to entropic effects.   

We have also demonstrated that coarse-grained modeling with Wang-Landau sampling can be a useful methodology for calculating $J$-factors for short DNA chains. This basic approach could find further application in exploring additional aspects of DNA bending elasticity, of great relevance for DNA compaction and protein-DNA interactions. Elaborations of the MWLC model may be required for these purposes. Our existing computational framework could easily incorporate the dependence of bending elasticity on salt concentrations and DNA sequence, through altered free energies of melting~\cite{SantaLuciaJr:1998} and altered flexibility of a given melted or hybridized section.~\cite{Baumann:1997p7135,Olson:1998p6734} Other possible straightforward amendments include more detailed melt thermodynamics (for example {\add incorporating stacking interactions,~\cite{Krueger:2006p2767}} a specific heat term \cite{Williams:2001p23892} or spatial and dynamical information~\cite{Dauxois:1993p2680}), volume exclusion, and explicit representation of DNA helicity.~\cite{Mastroianni:2009p46259} {\add Nevertheless, the meltable WLC still provides a parsimonious model lending important insights regarding the effect on DNA cyclization behavior of greater flexibility conferred by thermally-excited defects. }

\begin{acknowledgments}
Steve Whitelam provided many useful discussions.  DS acknowledges support from a National Science Foundation Graduate Research fellowship.  
\end{acknowledgments}

\end{document}